\documentclass[twocolumn,10 pt,showpacs,preprintnumbers,amsmath,amssymb]{revtex4}
\usepackage{amssymb}
\usepackage{graphicx}

\begin{document}

\title{Galvanomagnetic properties and noise in a barely metallic film of $V_2O_3$.}
\author{Clara Grygiel}
\altaffiliation{Current address: Department of Chemistry, The University of Liverpool, Liverpool L69 7ZD, United Kingdom.}
\author{Alain Pautrat}
\email{alain.pautrat@ensicaen.fr}
\affiliation{Laboratoire CRISMAT, UMR 6508 du CNRS, ENSICAEN et Universit\'{e} de Caen, 6 Bd Mar\'{e}chal Juin, 14050 Caen, France.}
\author{Pierre Rodi\`ere}
\affiliation{Institut N\'eel, CNRS-UJF, BP 166, 38042 Grenoble Cedex 9, France}

\begin{abstract}
We have measured the magnetotransport properties of a strained metallic $V_2O_3$ thin film.
Most of the properties are similar to $V_2O_3$ single crystals that have been submitted to a large pressure.
In addition, resistance noise analysis indicates that conductivity fluctuations are freezing out at $T\approx 10K$.
Examination of a range of measurements leads to the conclusion that spins-configuration fluctuations dominate in the low temperature regime.
\end{abstract}


\maketitle

\section{Introduction}
$V_2O_3$ single crystals undergo a first order metal-insulator (M-I) transition at $T\approx 160K$ with antiferromagnetic ordering of Vanadium spins associated to a structural transition \cite{V2O3}.
 In Vanadium deficient samples, the metallic phase was shown to be stabilized down to $T\approx 10K$ \cite{carter1},
 where a spin density wave (SDW) condenses as indicated by a clear increase of the resistivity \cite{bao}. At roughly the same temperature, the Hall resistance shows a maximum, which was first attributed to skew scattering in an ordered magnetic state \cite{rosenbaum}. 
 Stoichiometric $V_2O_3$ behaves differently. When submitted to a high hydrostatic pressure $P\geq 26 KBars$,
 its longitudinal resistivity decreases monotonically with a decrease in temperature and any gap opening can be observed.
 This indicates a fully suppressed metal to insulator
 transition down to the lowest temperature without any trace of magnetic ordering \cite{carter1}. On the other hand, the Hall resistance still shows
a maximum at $T\approx 10K$ \cite{klimm}, which may be explained by dominant spin fluctuations \cite{nari} or strong electronic correlations \cite{correlation} in the low temperature range.
 It has been reported that the spin susceptibility of metallic $V_2O_3$ under pressure exhibits also an anomaly at $T\approx 10K$ which has been associated with either low energy magnetic excitations or a kind of pseudo-gap formation \cite{suc}.
 There is a need for complementary experiments to clarify the origin of this anomaly.

 When $V_2O_3$ is epitaxially grown
  over a substrate, clamping of the film can prevent the structural transition \cite{clara}.
 As a consequence, a metallic state which mimics the highly pressured state is observed. Macroscopically, a distribution of strains
 in the thin film can lead to phase coexistence. The resistivity exhibits thermal hysteresis due to metastable states and is a non ergodic quantity
 in parts of the phase diagram, e.g. quantitative analysis of the measured values is ambiguous \cite{relax}.
 However, using a microbridge as a local probe, it is possible to isolate a pure metallic state without
 any thermal hysteresis and no trace of the M-I transition.
 In this paper, we discuss the transport properties of a strained $V_2O_3$ thin film in the so-called barely metallic side.
 We have performed conventional galvanomagnetic measurements and noise measurements. We will show that most of the properties are comparable
 to those observed in single crystals which were submitted to high pressure.
 The small size of the system makes the statistical averaging less effective
and, thus makes extraction of informations from noise measurements possible. As we will discuss below, the analysis of conductivity
 fluctuations provide new insights into the low temperature ground state of metallic $V_2O_3$.

 \section{Experimental}
The $V_2O_3$ thin film was grown on a substrate of (0001)-oriented saphire using
 the pulsed laser deposition from a $V_2O_5$ target. The details of the growth conditions,
 and some structural and microstructural characterisation have been reported previously \cite{clara}.
 In particular, the evolution of unit cell parameters was shown to be inconsistent with oxygen non stoichiometry \cite{clara}. 
The film studied here is a 230\,{\AA} thick sample, with a low rms roughness of 0.47\,{nm}
(averaged over 3$\times$3$\mu m^{2}$). It was patterned using UV Photolitography and argon ion etching
 to form a bridge of length$\times$width=$200\times100$ $\mu m^2$. 
Silver contact pads were connected
using aluminum-silicon wires (or Copper wires for measurements at temperature below 2K)
 which were attached by ultrasonic bonding for four probe and Hall effect measurements. 
Measurements for $T\leq 2 K$ were performed in a He-3 cryostat. Transport measurements were taken at
 temperatures between $2K$ and $400K$ in a Physical Properties Measurement System from Quantum Design (PPMS). All magnetotransport 
 measurements have been performed with the magnetic field along the c-axis, i.e. perpendicular to the film.
For noise measurements, we used a home-made sample holder and external electronics to acquire the resistance-time series and spectrum noise
 (the acquisition part consists of home-assembled low noise preamplifiers, a spectrum analyzer SR-760 and a Dynamic Signal Analyzer NI-4551).
 The final resolution is dominated by the low noise current supply (current noise: 0.2 $nA/Hz^{1/2}$)
 which leads to an equivalent noise of 100 $nV/Hz^{1/2}$ when the sample with $R\approx$ 0.5 $K\Omega$ is biased at low temperature. This is higher than both the equivalent Johnson thermal noise and the preamplifier noise.  
The length dependence of the noise measured between different arms of the bridge allow to conclude that contact noise is not important here \cite{contact}.

\section{Results}
\subsubsection{Magnetotransport properties}

Fig.1 shows the resistivity of the metallic microbridge as function of the temperature. The resistivity is thermally reversible
 and ohmic for applied currents up to at least 1 mA and from $T=400K$ down to $2K$. Both the functional form of the resistivity and the absolute
 value were close to what is measured in crystals under high pressure \cite{carter1}. In particular,
 the resistivity tends to curve downward at high temperature but takes a value above the 
3D Ioffe-Regel limit for metallic conductivity. This is often observed in so-called bad metals,
 and a large $\rho_{sat}\approx 1 m\Omega.cm$ has been discussed as a possible consequence of interacting
 electrons \cite{gun}.
For a correlated metal, a lot of different terms can play a role in electronic scattering,
 but at high temperature a major contribution from phonons can be still expected.
The electron-phonon
contribution to the resistivity is usually described by the Bloch-Gr$\ddot{u}$neisen formula:
 
\begin{equation}
\rho_{ph} = c.T^5/\theta6  \int_0^{\theta/T} x^5 [(e^x-1)(1-e^{-x})]^{-1} dx
\end{equation} 

where $\theta$ is the Debye temperature and $c$ is a constant describing the electron-phonon interaction.
 The high temperature data were adjusted to the phenomenological parallel resistor with a saturation resistivity acting as a shunt \cite{wiessman}.

\begin{equation}
\rho^{-1}=\rho_{sat}^{-1}+(\rho_{0}+\rho_{ph})^{-1}
\end{equation} 
 
A good fit is obtained for the temperature range $ 130K \leq T \leq 400K$. We found an approximative residual resistivity $\rho_{0}\approx 320 \mu \Omega .cm$ which will be refined below, a saturation resistivity $\rho_{sat}\approx 1 m\Omega.cm $, and a Debye temperature $\theta=300 K$.
 The deduced Debye temperature was close to half the value of bulk samples ($\theta_{bulk} \approx 560 K$),
 which could be qualitatively explained by a size effect \cite{debye} and/or by stress in the film.
For $T\leq 140-150K$, an additional contribution is needed to describe the resistivity.
 In particular, at the lowest temperatures, where the electron-phonon scattering was clearly negligible ($T\leq 20 K$),
 a quadratic temperature dependence of the resistivity was observed in the form of $\rho_{T^2}=A.T^2+\rho_0$, with $A=0.073 \mu \Omega.cm.k^{-2}$
 and a more precise value of the residual resistivity $\rho_{0}\approx 316 \mu \Omega .cm$(fig.2). 
 This Fermi liquid-like dependence has been reported
for single crystals of $V_2O_3$ submitted to a high pressure \cite{carter2,mederle}, and, more recently, in strained thin films \cite{clara2}.
 This dependence is typical
 of electron-electron scattering \cite{baber,lawrence} which is the general interpretation in $V_2O_3$. Spin fluctuations can also lead to a quadratic temperature dependence at low temperature \cite{nari}.
We note also that the functional form of $\rho(T)$ over the full temperature range was not very different from what was expected taken in spin fluctuations theories \cite{nari,coblin}.
 The $T^2$ variation
 of the resistivity did not extend down to the lowest temperature, as can be seen in fig.2.
 Here an excess of resistivity was observed when $T\leq  10K$.
After carefully subtracting $\rho_{T^2}$ from the measured resistivity at low temperature ($T\leq 20K$), a logarithmic
 dependence of the resistivity was observed from $T\leq 10K$ down to $2K$ (fig.3). This is the limiting temperature in our PPMS cryostat.
 Measurements in another film of the same thickness at the lowest
 temperature indicated that this logarithmic dependence extends down to $400 mK$. Kondo effect (scattering by magnetic impurities)
 or weak localisation corrections can be made to explain that the scattering increases at low temperatures with such characteristics. 
The
 $T^{1/2}$ dependent conductivity which is characteristic of electron-electron scattering in a disordered medium has been proposed 
as an explanation
 for the low temperature upturn in pressurized $V_2O_3$ resistivity \cite{carter1,mederle}. This however does
 not provide a satisfactory fit for our sample.
 
In addition to the excess of resistivity, a negative longitudinal magnetoresistance (MR) appears for $T\leq 130K$ (inset of fig.4).
 Its magnetic field dependence is quadratic from $B=0T$ to $7T$ and for all temperatures down to $2K$. The origin of the 
quadratic dependence 
of a negative magnetoresistance is uncertain \cite{griessen}. One possible mechanism is 2D weak localization (WL) as far as $B \ll  \hbar /(4eL_\varphi^2)$, where $L_\varphi$
 is the coherence length \cite{notebene}.
 Quadratic negative magnetoresistance is also a common feature of Kondo-like magnetic metals, where the magnetic field tends to suppress the spin fluctuations.
 Remarkably, the absolute value of the MR shows a notable increase for $T\leq 10K$, clearly
evidenced by tracing the Kohler's plot. The idea behind Kohler's rule is that in conventional isotropic metals,
 the magnitude of MR is fixed by a single scattering time $\tau(T)\propto 1/\rho(T)$, and implies 
that $\Delta \rho(B)/ \rho(0)=F(B \tau)$ 
where $F$ is a function dependent on the details of the electronic structure. Due to its quite large generality, 
 the Kohler's rule is expected to apply in a Fermi liquid also and has been used as a probe of non Fermi liquid ground states. 
In the limit of the carrier density being temperature independent \cite{luo},
 Kohler's rule written in its simplest form is $\Delta \rho(B) / \rho(0)=F(B / \rho(0))$.
 In fig.4, it can be seen that $\Delta\rho(B)/\rho(0)$ plotted as function of $(B/\rho(0))^2$ for different
 temperatures gives a single curve, i.e. Kohler's rule is fulfilled when $T\geq 10K$. The strong increase of the MR below $10K$
 was thus associated with a departure from Kohler's rule. 
The Hall resistance also presented a maximum
 in the 10K-20K range (fig.5), as previously reported both for single crystals and for thin films \cite{rosenbaum,klimm,clara2}.
Since it was observed in absence of a clear magnetic transition, a dominant effect of spin-fluctuation has been proposed \cite{nari,klimm}.
 Such a maximum can also be expected from strong correlation effects \cite{correlation}.
In summary of this part, we have confirmed that the magnetotransport properties of metallic $V_2O_3$ thin film present similarities with $V_2O_3$ crystals under pressure
 such as the $T^2$ variation of the resistivity and the maximum of the Hall effect at $T\approx 10K$.
 In addition, we have shown that the $10K$ anomaly reflects also in the MR and in a Kondo-like increase of the resistivity at low temperature.
Both electronic correlations and spin fluctuations can be used as qualitative explanations. 

\subsubsection{Noise measurements as a complementary tool}

Benefiting from the small size of our sample, we have studied conductivity fluctuations
 by measuring the resistance noise. We have measured a set of resistance-time series
 at various temperatures. The temperature was allowed to stabilize during sample cooling before
 each measurement series.
 The power spectral density (or noise spectrum) $S_{RR}(f)$ is then calculated. 
 In ordinary metals,
 the electronic noise generally comes from the motion of atomic impurities or defects and the temperature
 dependence is dependent by the underlying mechanisms, e.g. thermal activation or eventually tunneling at low temperature. 
 At high temperature $T>100K$, we have observed resistance switches insensitive to magnetic field
 during long periods, which recall the Hydrogen
 hopping noise common in several metals \cite{mikehydrogen}. The fractional change in resistance was between
 $2.10^{-5}-10^{-4}$, which is large for individual hopping and implied a collective motion.
  Note that this switching noise strongly evolved over a long period, as it was not observed one month after the first measurements,
 showing that the involved defects were mostly non-equilibrium defects left in the film after deposition.
 In addition, reproducible $1/f$ noise can be observed
 and this noise will be discussed now. 
 
 Shown in fig.6 is typical noise spectra measured on our film.
 They are very close to $1/f$ over the whole temperature range.
The temperature dependence of this normalized noise exhibits a non monotonic variation (fig.7),
 with a plateau in the range $10K<T<20K$. The sudden decrease of the noise at $T\approx 10K$ is remarkable.
To give an order of magnitude of the noise level,
 we use the phenomenological Hooge parameter
 which is, for $1/f$ noise, $\gamma=(S_{RR}(f).f)/R^2.n_c.V$ where $n_c$ is the carriers density and $V$ the (probed) sample volume \cite{black}.
  For homogeneous and pure metals, typical values
 are in the range $\gamma\approx 10^{-2}-10^{-3}$. Taking $n_e=5.10^{22} cm^{-3}$ deduced from our Hall effect measurements,
 we found $\gamma\approx 10^{-1}$ at room temperature, which is large, but not unusually large for a metallic-like thin
 film and is consistent with a moderately higher concentration of defects in films than in bulk. 
 Noise in metals can be usually explained with Duttah-Dimon-Horn (DDH) analysis, which assumes that the electronic fluctuations
 come from smeared kinetics
 of defects with a distribution of activation energies \cite{dutta}. When the defect relaxation is activated and typical energy
 is in the $eV$ range, the noise generally increases with temperature. However, there is at least one report of a noise maximum at low temperature which however follows the DDH framework.
 It has been interpreted as evidence of low energy excitations in an oxide \cite{ghosh}.
In the DDH modeling, there is a direct relation between the noise spectral exponent $\alpha$ and
 the temperature dependence of the noise in the form $\alpha(T)=1-(\partial{ln S_{RR}}/\partial{ln T}-1)/ln(2\pi f \tau_0)$,
 where $\tau_0^{-1}$ is the attempt frequency.
We have checked this relation and found that unrealistically large $\tau_0$ values are needed to approach
 the very small variation of $\alpha(T)$. In parallel, the temperature of the noise maximum/plateau does not show any frequency dependence.
 Actually, pure 1/f noise should follow $S_{RR}\propto T$ in the DDH model, as we observed at high temperature (fig.7), and a maximum is not expected.
 We conclude that the low temperature noise maximum does not come from thermally activated excitations. It can neither be explained
 by temperature fluctuations which would lead to a too weak temperature dependence of the noise.
 In disordered metals at low temperature, universal conductance fluctuations (UCF) can
 in principle take place \cite{UCF}. In the UCF model, the noise grows at low temperature due to the enhancement of the effective coupling between disorder configurations and resistance changes.
UCF can explain why $S_{RR}$ rises when the temperature decreases, but not the apparent freezing of conductivity fluctuations for $T\leq 10K$.
 Since we observe a decrease of the resistance noise when a magnetic field is applied in the $10 K-30 K$ range
 (a factor of 0.6 at B$=7T$ and at T$=15K$, see the inset of fig.7), a spin-coupling origin of this noise can be suggested.
  The blocking of magnetic fluctuations at low temperature with non Arrhenius slowing down makes think of spin glasses \cite{MB}.
 Note, however, that some spin glasses do not show a decrease but a growing of noise at low temperature due to UCF coupling \cite{CuMn}.  
 To probe, directly, spins freezing with resistance noise, the sensitivity of the resistance to spins should be independent 
of the temperature and the coupling should be local \cite{CuMn}.
 This appears to be the case with our sample. Note that in some intermetallic compounds, similar properties, such as a quadratic 
temperature dependence 
of the resistivity close to a peak in the Hall coefficient have been interpreted as a spin glass type of freezing 
at low temperature \cite{spinglass,spinglass2}. A precise determination of the magnetic disorder is required
 to better understand its role in the low temperature galvanomagnetic properties of $V_2O_3$.

To conclude, we have measured a large array of transport properties in a $V_2O_3$ film in the barely metallic side.
 We found strong similarities with properties of single crystals submitted to a high pressure, 
 and some new features such as a low temperature noise maximum coupled to anomalous magnetotransport properties.
 A central role of spin-configuration fluctuations, which freeze at low temperature, has been proposed. Future experiments 
directly sensitive to magnetic order would be extremely complementary. Noise experiments on single crystals, if feasible,
 would be, also, very interesting to compare with our measurements in films.
 
 Acknowledgements: A.P. thanks W. Prellier for comments on the manuscript and C. G. thanks S. McMitchell.

\newpage
\begin{figure}[t!]
\begin{center}
\includegraphics*[width=8.0cm]{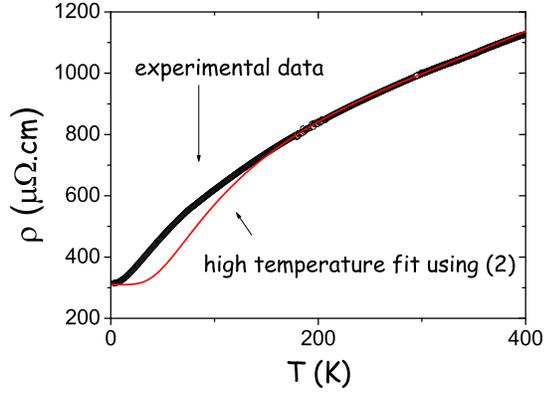}
\end{center}
\caption{Resistivity of the metallic $V_2O_3$ microbridge as function of temperature. Also shown is a fit of the high temperature part with the formula (2) ($\theta=300K$, $\rho_{sat}= 1 m\Omega.cm$).}
\label{fig.1}
\end{figure}

\begin{figure}[t!]
\begin{center}
\includegraphics*[width=8.0cm]{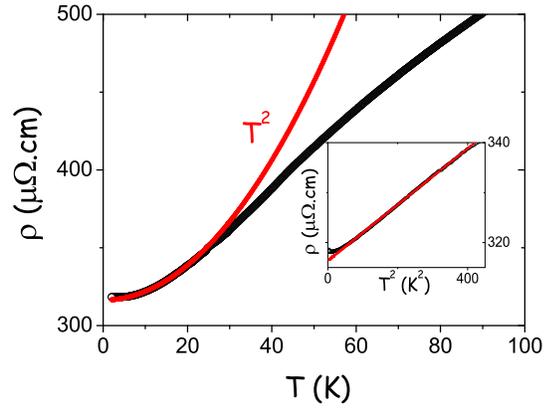}
\end{center}
\caption{Resistivity of the metallic $V_2O_3$ microbridge as function of temperature for $T\leq 100K$ and $T^2$ fit of the low temperature part.
Shown in the inset is a zoom of $\rho$ as function of $T^2$. The temperature variation of the resistivity is quadratic for $10K\leq T\leq 20K$.}
\label{fig.2}
\end{figure}

\begin{figure}[t!]
\begin{center}
\includegraphics*[width=8.0cm]{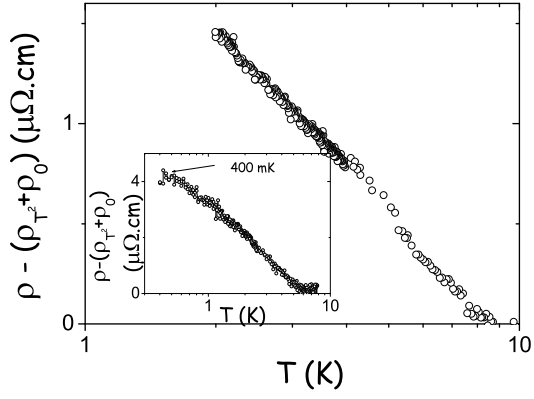}
\end{center}
\caption{Variation of the resistivity corrected from the quadratic temperature dependence and from the residual resistance.
 A logarithm dependence can be observed for $2K\leq $T$\leq 10K$. Shown in the inset is the same characteristics in a film of the same thickness and measured down to lowest temperature ($400 mK$).}
\label{fig.3}
\end{figure}

\begin{figure}[t!]
\begin{center}
\includegraphics*[width=8.0cm]{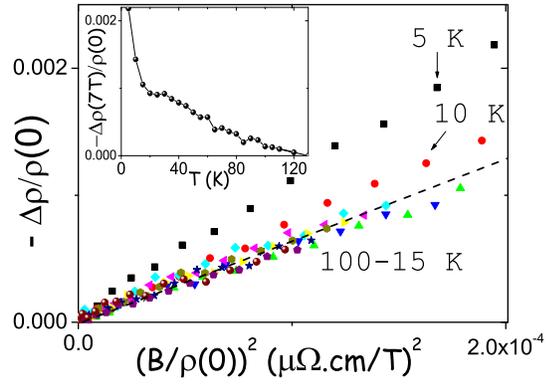}
\end{center}
\caption{A Kohler's plot of the magnetoresistance $\Delta\rho(B)/\rho(0)$ as function of $(B/\rho(0))^2$ for temperature from 100K to 5K. The dashed line is a guide for the eyes.
 Shown in the inset is the magnetoresistance at 7T as function of the temperature.}
\label{fig.4}
\end{figure}

\begin{figure}[t!]
\begin{center}
\includegraphics*[width=8.0cm]{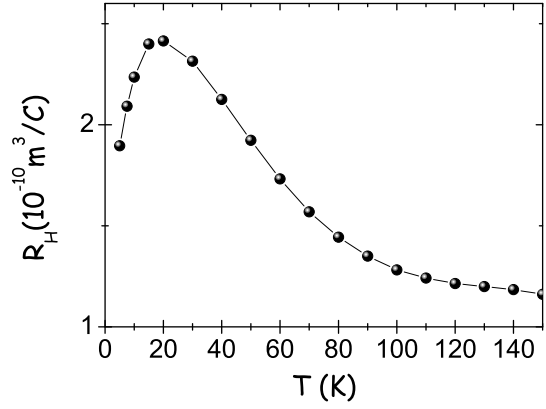}
\end{center}
\caption{Hall resistance as function of the temperature, showing the maximum at low temperature.}
\label{fig.5}
\end{figure}

\begin{figure}[t!]
\begin{center}
\includegraphics*[width=8.0cm]{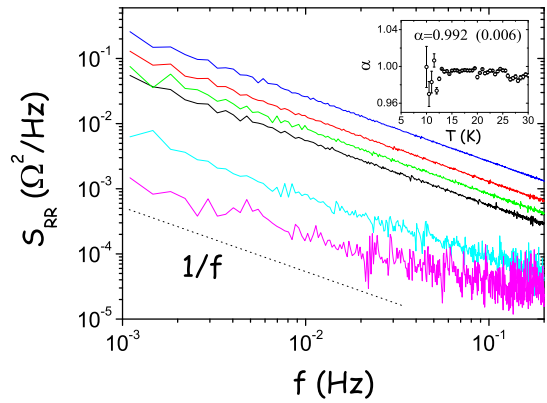}
\end{center}
\caption{$1/f^{\alpha}$ noise spectra obtained from the resistance time series at T$=5K, 10K, 30K, 25K, 20K, 15K$ from the bottom 
to the top. Shown in the inset is the value of the exponent $\alpha$ as function of the temperature.}
\label{fig.6}
\end{figure}

\begin{figure}[t!]
\begin{center}
\includegraphics*[width=8.0cm]{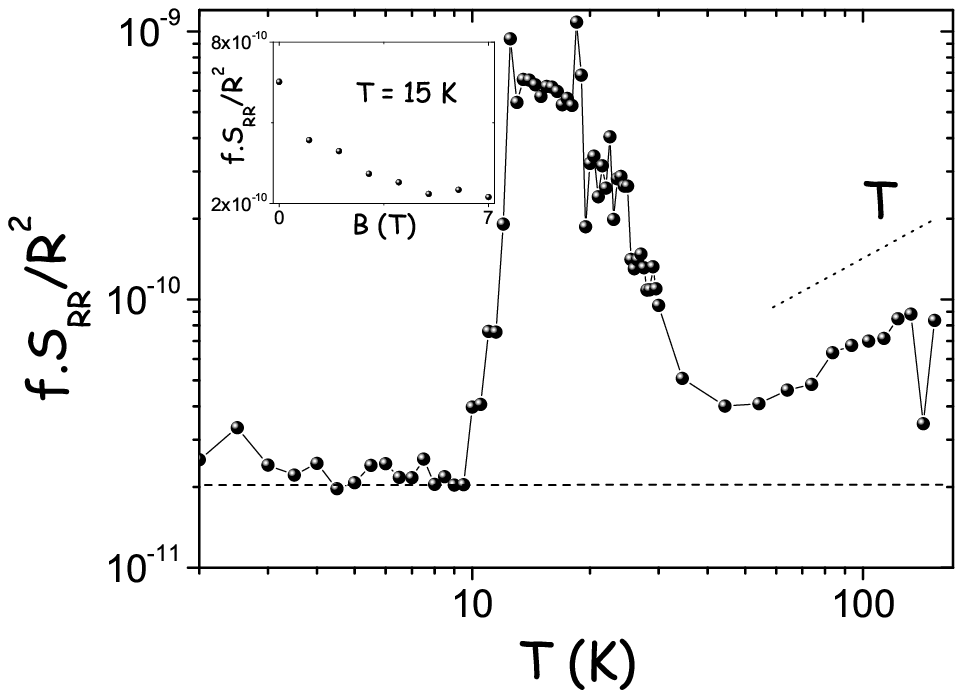}
\end{center}
\caption{Normalized resistance noise as function of the logarithm of the temperature (f$=0.1 Hz$). Note the sharp decrease at T$\approx 10K$. The horizontal dotted line corresponds to the resolution of the measurement.
 Shown in the inset is the normalized noise as function of the magnetic field for $T=15K$, (f$=0.1 Hz$).}
\label{fig.7}
\end{figure}


\begin{references}
\bibitem{notebene} We deduce a higher bound $L_\varphi\leq  5 nm$. This low value of the coherence length could be attributed to some defects due to the strains in the film at low temperature, rather than to a low carriers
 density $n_e$, because Hall effect measurements give $n_e\approx 4.10^{22}/cm3$ at room temperature.
\end{references}
\end{document}